\documentclass[prb,twocolumn]{revtex4}%

\usepackage{graphicx}
\usepackage{amsmath}
\usepackage{ulem}
\usepackage{color}

\newcommand{\be}{\begin{equation}}
\newcommand{\ee}{\end{equation}}
\newcommand{\bea}{\begin{eqnarray*}}
\newcommand{\eea}{\end{eqnarray*}}
\newcommand{\bean}{\begin{eqnarray}}
\newcommand{\eean}{\end{eqnarray}}

\begin{document}

\draft
\title{\bf Thermoelectric Properties of Armchair Graphene Nanoribbons with Array Characteristics}

\author{David M T Kuo}

\address{Department of Electrical Engineering and Department of Physics, National Central
University, Chungli, 32001 Taiwan}

\date{\today}

\begin{abstract}
The thermoelectric properties of armchair graphene nanoribbons
(AGNRs) with array characteristics are investigated theoretically
using the tight-binding model and Green's function technique. The
AGNR structures with array characteristics are created by
embedding a narrow boron nitride nanoribbon (BNNR) into a wider
AGNR, resulting in two narrow AGNRs. This system is denoted as
w-AGNR/n-BNNR, where 'w' and 'n' represent the widths of the wider
AGNR and narrow BNNR, respectively. We elucidate the coupling
effect between two narrow symmetrical AGNRs on the electronic
structure of w-AGNR/n-BNNR. A notable discovery is that the power
factor of the 15-AGNR/5-BNNR with the minimum width surpasses the
quantum limitation of power factor for 1D ideal systems. The
energy level degeneracy observed in the first subbands of
w-AGNR/n-BNNR structures proves to be highly advantageous in
enhancing the electrical power outputs of graphene nanoribbon
devices.
\end{abstract}

\maketitle

\section{Introduction}
Extensive research efforts have been dedicated to exploring the
potential applications of graphene nanoribbons (GNRs) across
various fields such as electronics, optoelectronics, and
thermoelectric devices. This interest has surged since the
groundbreaking discovery of two-dimensional graphene in 2004 by
Novoselov and Geim [\onlinecite{Novoselovks}]. Despite significant
strides, GNR-based devices face a pronounced challenge in
amplifying their electrical and optical power outputs. The limited
power outputs are attributed to the low transmission coefficient
in the band edges of the first subbands of GNRs
[\onlinecite{JustinH}--\onlinecite{WangHM}]. In the context of
GNR-based device applications, the electronic states proximate to
the band edges of the initial conduction and valence subbands play
a pivotal role in optical and transport processes. Consequently,
it becomes imperative to engineer band-edge electronic states with
a high transmission coefficient, paving the way for the
development of electronic and thermoelectric devices capable of
enhancing their electrical power outputs.

Different classes of GNRs have undergone thorough theoretical and
experimental investigations by diverse research groups. These
encompass armchair GNRs (AGNRs)
[\onlinecite{Nakada},\onlinecite{Cai}], zigzag GNRs (ZGNRs)
[\onlinecite{Wakabayashi}], cove-edged zigzag GNRs (CZGNRs)
[\onlinecite{LiuJ}--\onlinecite{WangX}], AGNR heterojunctions
[\onlinecite{ChenYC}--\onlinecite{Kuo3}], and graphene quantum dot
superlattices [\onlinecite{Sevincli},\onlinecite{YangK}].
Typically, these quasi-one-dimensional systems exhibit suboptimal
transmission coefficients near the edge states of the first
conduction and valence subbands when coupled with electrodes
[\onlinecite{Areshkin}--\onlinecite{Kuo1}]. In the realm of
ballistic transport, under ideal conditions and neglecting
electron spin degeneracy, the one-dimensional transmission
coefficient for electrons in the first conduction and valence
subbands is anticipated to be unity. However, defects and contact
effects in finite Graphene Nanoribbons (GNRs) inevitably introduce
backward scattering, reducing the transmission coefficient to less
than one for specific electron wavelengths near the band edge of
the first subbands [\onlinecite{Areshkin}]--[\onlinecite{Kuo1}].
Additionally, the contact geometries between the graphene
electrodes and the molecules play a significant role in
influencing the transmission coefficient of electron transport in
the molecules [\onlinecite{LiX}]--[\onlinecite{Liyun}].
Consequently, achieving an energy-dependent transmission
coefficient of one for electrons across all wavelengths becomes
challenging when finite GNRs are connected to electrodes. It was
demonstrated that in the case of quantum dot (QD) molecules with
high orbital degeneracy, a greater degree of degeneracy leads to a
higher transmission coefficient. This increase in transmission
coefficient enhances electrical conductance while keeping the
Seebeck coefficient unchanged. Consequently, such enhancement
results in increased electrical power outputs in QD-based
thermoelectric devices [\onlinecite{DavidK}].

In this study we propose an innovative configuration wherein a
narrow boron nitride nanoribbon (BNNR) is seamlessly integrated
into a wider AGNR. The realization of such a structure can be
achieved through advanced DUV lithographic techniques
[\onlinecite{LiuZ}]. We explore AGNRs and BNNRs with varying
widths, spanning from 7 to 19 for AGNRs and 3 to 9 for BNNRs.
These configurations are denoted as w-AGNR/n-BNNR, illustrating
scenarios where electrons transport along the armchair direction.
Figures 1(a)-1(f) portray six instances of w-AGNR/n-BNNR
structures, which can alternatively be interpreted as
u-AGNR/m-BNNR/b-AGNR heterostructures, where u, m, and b represent
the widths of the upper AGNR, middle BNNR, and bottom AGNR,
respectively. A noteworthy discovery from our study is that the
power factor of the 15-AGNR/5-BNNR, forming a two narrow
symmetrical AGNR array, surpasses the quantum limitation of the
power factor for one-dimensional ideal systems. The orbital
degeneracy in the first subbands of 15-AGNR/5-BNNR, characterized
by a substantial band gap, significantly enhances electrical
conductance. Meanwhile, the semiconducting phase of two narrow
AGNRs maintains an unchanged Seebeck coefficient. Consequently,
the power factor of the 15-AGNR/5-BNNR junction is greatly
enhanced.

\begin{figure}[h]
\centering
\includegraphics[trim=1.cm 0cm 1.cm 0cm,clip,angle=0,scale=0.3]{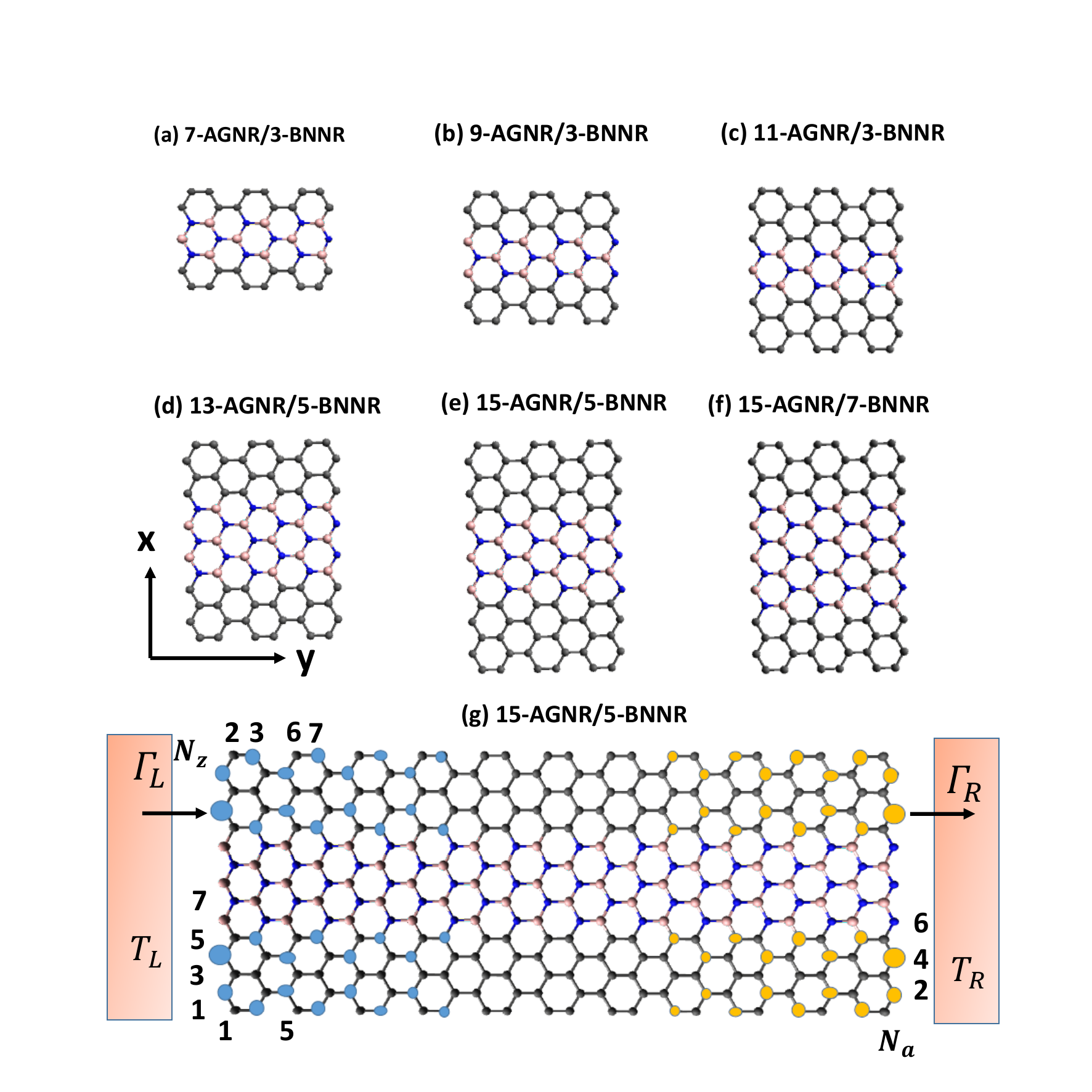}
\caption{Schematic diagram depicts hybridized w-AGNR/n-BNNR
structures, where the notations w and n refer to the wide and
narrow widths, respectively. Panels (a)-(f) showcase six different
scenarios, while panel (g) illustrates the line-contacting of
zigzag-edge atoms in a 15-AGNR/5-BNNR structure to electrodes.
Symbols $\Gamma_{L}$ ($\Gamma_R$) represent the electron tunneling
rate between the left (right) electrode and the leftmost
(rightmost) atoms at the zigzag edges, and $T_{L}$ ($T_{R}$)
denotes the equilibrium temperature of the left (right) electrode.
The charge densities of $\varepsilon_{e,1} = 0.3741$ eV and
$\varepsilon_{h,1}=-0.3932$ eV for 15-AGNR/5-BNNR structure are
depicted in panel (g), with light-blue and orange circles
representing the charge densities for $\varepsilon_{e,1}$ and
$\varepsilon_{h,1}$, respectively. The radius of the circle
represents the intensity of the charge density.}
\end{figure}

\section{Calculation Methodology}
To explore the thermoelectric properties of w-AGNR/n-BNNR
connected to the electrodes, we utilize a combination of the
tight-binding model and the Green's function technique. The system
Hamiltonian consists of two components: $H = H_0 + H_{GNR}$. Here,
$H_0$ signifies the Hamiltonian of the electrodes, encompassing
the interaction between the electrodes and the w-AGNR/n-BNNR.
Meanwhile, $H_{GNR}$ represents the Hamiltonian for the
w-AGNR/n-BNNR and can be expressed as follows:

\begin{small}
\begin{eqnarray}
H_{GNR}&= &\sum_{\ell,j} E_{\ell,j} d^{\dagger}_{\ell,j}d_{\ell,j}\\
\nonumber&-& \sum_{\ell,j}\sum_{\ell',j'} t_{(\ell,j),(\ell', j')}
d^{\dagger}_{\ell,j} d_{\ell',j'} + h.c,
\end{eqnarray}
\end{small}

Here, $E_{\ell,j}$ represents the on-site energy of the orbital in
the ${\ell}$-th row and $j$-th column. The operators
$d^{\dagger}_{\ell,j}$ and $d_{\ell,j}$ create and annihilate an
electron at the atom site denoted by ($\ell$,$j$). The parameter
$t_{(\ell,j),(\ell', j')}$ characterizes the electron hopping
energy from site ($\ell'$,$j'$) to site ($\ell$,$j$). We assign
the tight-binding parameters for w-AGNR/n-BNNR as follows: $E_{B}
= 2.329$ eV, $E_{N} = -2.499$ eV, and $E_{C} = 0$ eV to boron,
nitride, and carbon atoms, respectively. To simplify our analysis,
we have neglected variations in electron hopping strengths between
different atoms due to their relatively minor differences
[\onlinecite{GSSeal}]. We set $t_{(\ell,j),(\ell',j')} = t_{pp\pi}
= 2.7$ eV for the nearest-neighbor hopping strength.  We can
utilize these parameters to replicate the bandgaps of the
BNNR/AGNR/BNNR structure as illustrated in Fig. 3(d) of reference
[\onlinecite{DingY}], which were originally calculated using the
first-principle method.

In the linear response region, the electrical conductance ($G_e$),
Seebeck coefficient ($S$) and  electron thermal conductance
($\kappa_e$) can be computed using $G_e=e^2{\cal L}_{0}$,
$S=-{\cal L}_{1}/(eT{\cal L}_{0})$ and $\kappa_e=
\frac{1}{T}({\cal L}_2-\frac{{\cal L}^2_1}{{\cal L}_0})$ with
${\cal L}_n$ ($n=0,1,2$) defined as

\begin{equation}
{\cal L}_n=\frac{2}{h}\int d\varepsilon~ {\cal
T}_{LR}(\varepsilon)(\varepsilon-\mu)^n\frac{\partial
f(\varepsilon)}{\partial \mu}.
\end{equation}

Here, $f(\varepsilon) = 1/(1+\exp((\varepsilon-\mu)/k_BT))$
represents the Fermi distribution function of electrodes at
equilibrium chemical potential $\mu$. The constants $e$, $h$,
$k_B$, and $T$ denote the electron charge, Planck's constant,
Boltzmann's constant, and the equilibrium temperature of the
electrodes, respectively. ${\cal T}_{LR}(\varepsilon)$ signifies
the transmission coefficient of a w-AGNR/n-BNNR connected to
electrodes, and it can be calculated using the formula ${\cal
T}_{LR}(\varepsilon) =
4Tr[\Gamma_{L}(\varepsilon)G^{r}(\varepsilon)\Gamma_{R}(\varepsilon)G^{a}(\varepsilon)]$
[\onlinecite{Kuo1}], where $\Gamma_{L}(\varepsilon)$ and
$\Gamma_{R}(\varepsilon)$ denote the tunneling rate (in energy
units) at the left and right leads, respectively, and
${G}^{r}(\varepsilon)$ and ${G}^{a}(\varepsilon)$ are the retarded
and advanced Green's functions of the GNRs, respectively. The
tunneling rates are determined by the imaginary part of the
self-energy originating from the coupling between the left (right)
electrode and its adjacent GNR atoms. In terms of tight-binding
orbitals, $\Gamma_{\alpha}(\varepsilon)$ and Green's functions are
matrices. For simplicity, $\Gamma_{\alpha}(\varepsilon)$ for
interface atoms possesses diagonal entries with a common value of
$\Gamma_t$ [\onlinecite{Kuo1}]. When graphene is connected to
metal electrodes, contact properties such as the Schottky barrier
or ohmic contact can exert a substantial impact on electron
transport in graphene [\onlinecite{Matsuda}]. Despite numerous
theoretical studies striving to elucidate this crucial behavior
from first principles, the theoretical limitations result in
obtaining only qualitative results regarding $\Gamma_t$ arising
from the contact junction [\onlinecite{LeeG}]. The thermoelectric
figure of merit is calculated by $ZT = S^2 G_e
T/(\kappa_e+\kappa_{ph})$, where $\kappa_{ph}$ is the phonon
thermal conductance of GNRs.

\section{Results and discussion}
\subsection{Electronic Structures of w-AGNR/n-BNNR Structures}
The electronic behavior of AGNRs is primarily determined by their
widths, which adhere to the rule $N_z = 3p$, $N_z = 3p + 1$, and
$N_z = 3p + 2$, where p is an integer. Specifically, AGNRs exhibit
semiconducting behavior for $N_z = 3p$ and $N_z = 3p + 1$, while
AGNRs with $N_z = 3p + 2$ exhibit either metallic behavior or
possess small band gaps in their electronic structures
[\onlinecite{SonYW}]. To illustrate the impact of BNNRs, the
electronic structures of various w-AGNR/n-BNNR structures
(u-AGNR/m-BNNR/b-AGNR heterostructures) are presented in Figure
2(a)-2(f).

As depicted in Figures 2(a)-2(f), w-AGNR/n-BNNR structures
distinctly exhibit semiconducting phases. However, for widths w $
= 7$, $9$, and $11$, we observe the absence of degeneracy in the
first subbands, attributed to the narrow barrier width between
u-AGNR and b-AGNR, which arises from BNNR with $ n = 3 $.
Insufficient width of 3-BNNR indicates coupling between two narrow
symmetrical AGNRs. This coupling effect induces bonding and
antibonding energy levels, lifting orbital degeneracy. The lack of
orbital degeneracy indicates a w-AGNR/3-BNNR without the
characteristic of an AGNR array. On the other hand, the degeneracy
of the first subbands exists in the case of the 15-AGNR/5-BNNR and
15-AGNR/7-BNNR structures. In these situations, w-AGNR/n-BNNR
structures exhibit AGNR array characteristics. It is worth noting
that the 5-AGNRs within the 15-AGNR/5-BNNR structure exhibit
semiconducting phases with a band gap of $E_{gap} = 0.7$ eV,
contradicting the characteristic metallic phase associated with
AGNRs of width $N_z = 5$ [\onlinecite{Wakabayashi}].

The original small band gaps of 5-AGNRs are enlarged due to a
change in one of the boundary conditions, transitioning from the
vacuum potential barrier to the potential barrier of BNNR. This
outcome aligns with the results predicted by the first principle
method [\onlinecite{DingY}], where the authors considered 5-AGNRs
confined by two BNNRs. By artificially setting the energy levels
of nitride and boron atoms to a large value, resembling a vacancy,
the 5-AGNRs revert to metallic phases.

\begin{figure}[h]
\centering
\includegraphics[angle=0,scale=0.3]{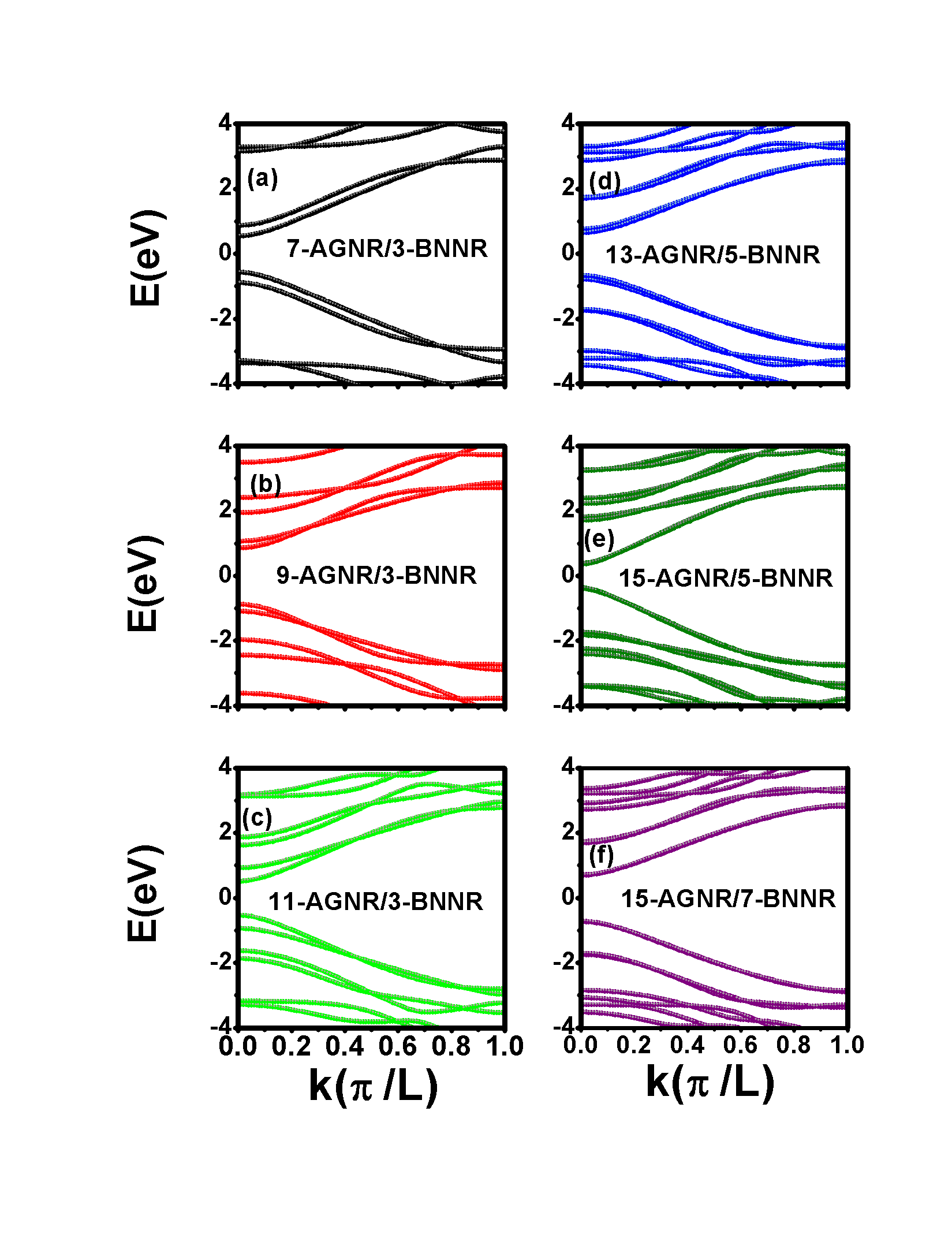}
\caption{Electronic subband structures of various w-AGNR/n-BNNR
structures: (a) 7-AGNR/3-BNNR, (b) 9-AGNR/3-BNNR, (c)
11-AGNR/3-BNNR, (d) 13-AGNR/5-BNNR, (e) 15-AGNR/5-BNNR, and (f)
15-AGNR/7-BNNR.}
\end{figure}

\subsection{Finite w-AGNR/n-BNNR Structures}
In the fabrication of GNR-based devices, ensuring that the channel
length is smaller than the electron mean free path is a crucial
requirement for achieving ballistic transport. To explore the
influence of channel length and contacts, we compute the energy
levels of 15-AGNR/5-BNNR structures that are decoupled from the
electrodes. The calculated results are presented in Fig. 3(a). The
energy levels, labeled as $\varepsilon_{e,1} = 0.3741$ eV and
$\varepsilon_{h,1} = -0.3932$ eV, are found to be relatively
insensitive to variations in $N_a$ within the range of 44 to 100.
However, the energy level separation $\Delta_e$ ($\Delta_h$)
between $\varepsilon_{e,1}$ ($\varepsilon_{h,1}$) and
$\varepsilon_{e,2}$ ($\varepsilon_{h,2}$) diminishes as $N_a$ is
increased.

The charge densities for $\varepsilon_{e,1} = 0.3741$ eV and
$\varepsilon_{h,1} = - 0.3932$ eV are depicted in Fig. 1(g), with
light-blue and orange circles representing the charge densities
for $\varepsilon_{e,1}$ and $\varepsilon_{h,1}$, respectively.
These charge densities exhibit a decay along the armchair
directions. Based on the distribution of charge density, it
becomes evident that $\varepsilon_{e,1}$ and $\varepsilon_{h,1}$
correspond to the end zigzag edge states of finite AGNRs. The
presence of BNNRs causes these energy levels of end zigzag edge
states in AGNRs to shift from zero energy modes to
$\varepsilon_{e,1}$ ($\varepsilon_{h,1}$). For energy values
within the range $0 < \varepsilon < 1$ eV, there are six energy
levels in the case of $N_a = 44$. As depicted in Figs. 3(b)-3(e),
the calculated transmission coefficient ${\cal
T}_{LR}(\varepsilon)$ for the 15-AGNR/5-BNNR structure with $N_a =
44$ ($L_a=4.54$ nm) clearly reveals these six energy levels only
for a small tunneling rate $\Gamma_t = 0.09$ eV.

In the case of a large tunneling rate, $\Gamma_t = 2.7$ eV, as
depicted in Fig. 3(e), which can be considered as resembling
graphene electrodes, the maximum values of ${\cal
T}_{LR}(\varepsilon)$ reach two in the first conduction subband.
This can be regarded as evidence showcasing 15-AGNR/5-BNNR with
AGNR array characteristic. It is important to note that the area
under the ${\cal T}_{LR}(\varepsilon)$ curve for $\Gamma_t = 2.7$
eV is maximized. Henceforth, we focus on the case with $\Gamma_t =
2.7$ eV throughout this article.

\begin{figure}[h]
\centering
\includegraphics[angle=0,scale=0.3]{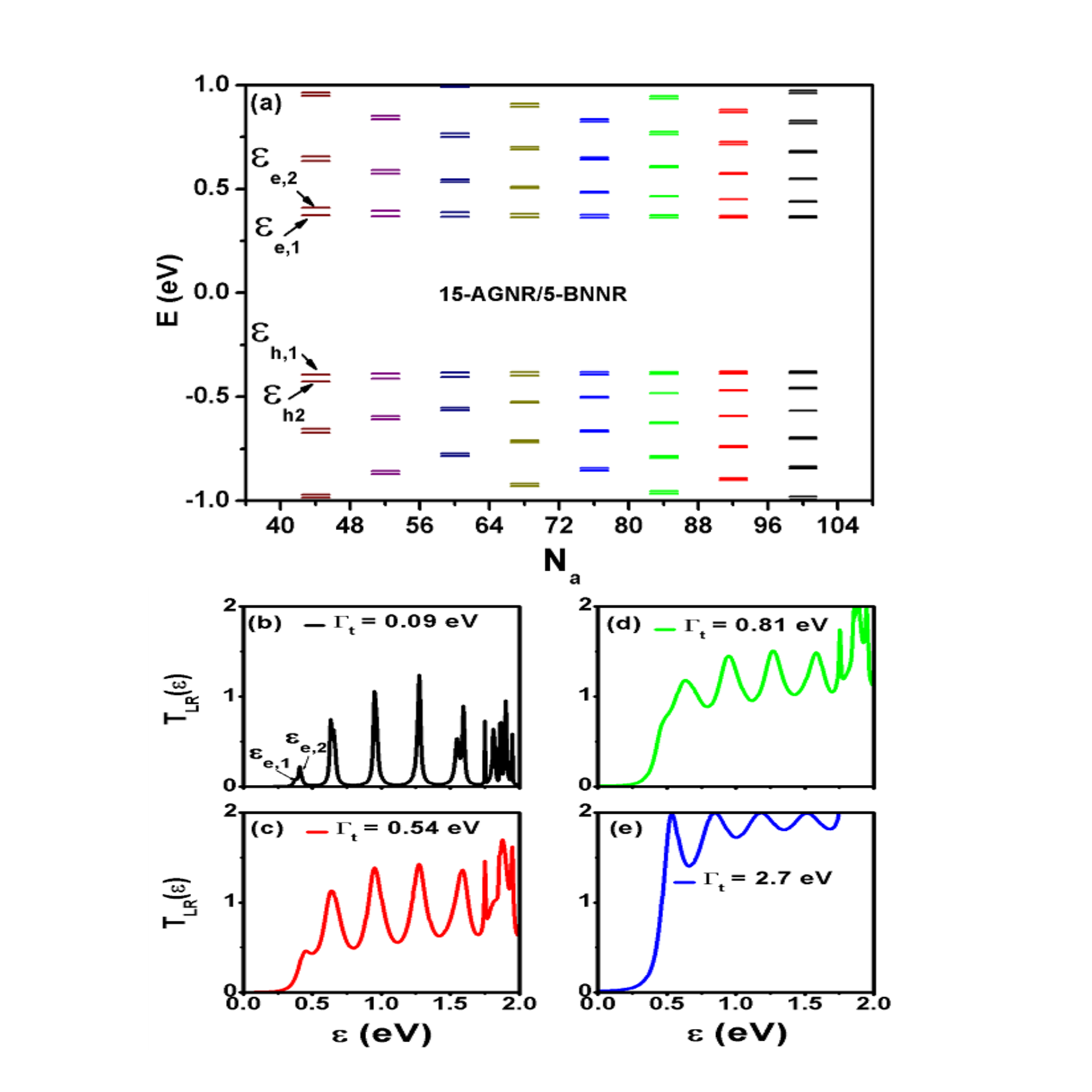}
\caption{Energy levels of finite 15-AGNR/5-BNNR structures for
various $N_a$ values (a). Transmission coefficient ${\cal
T}_{LR}(\varepsilon)$ of 15-AGNR/5-BNNR with $N_a = 44$ ($L_a =
4.54$ nm) for different $\Gamma_t$ values. (b) $\Gamma_t = 0.09$
eV, (c) $\Gamma_t = 0.54$ eV, (d) $\Gamma_t = 0.81$ eV, and (e)
$\Gamma_t = 2.7$ eV.}
\end{figure}

To investigate the impact of BNNRs on the transmission
coefficients of AGNR structures, we present the calculated ${\cal
T}_{LR}(\varepsilon)$ for four different w-AGNR/n-BNNR structures,
all characterized by $N_a = 84$ ($L_a = 8.8$ nm) and $\Gamma_t =
2.7$ eV in Fig. 4. In Figure 4(a), we observe that the 11-AGNR
transitions from a metallic phase to a semiconducting phase when
3-BNNR is embedded within it. However, the maximum values of
${\cal T}_{LR}(\varepsilon)$ in the first conduction and valence
subbands only reach one. As the second subbands emerge around
$\varepsilon \approx 1$ eV, ${\cal T}_{LR}(\varepsilon)$ can
attain a value of two. This behavior can be understood by
referencing the electronic structure in Figure 2(c).

In Figure 4(b), we observe that the band gap of the semiconducting
13-AGNR is widened when 5-BNNR is integrated into the 13-AGNR
structure. Moreover, its maximum ${\cal T}_{LR}(\varepsilon)$ can
reach two within the first subbands. However, the occurrence of
${\cal T}_{LR}(\varepsilon)$ equal to two is confined to specific
energy ranges. Figure 4(c) demonstrates that the transmission
coefficient curve of the 15-AGNR/5-BNNR exhibits a larger area
within the first conduction and valence subbands compared to the
1D-ideal case with a rectangular shape. Conversely, in the case of
7-BNNR, as shown in Figure 4(d), the area of ${\cal
T}_{LR}(\varepsilon)$ is reduced when compared to the 5-BNNR
scenario. It's worth noting that the 15-AGNR/7-BNNR, corresponding
to the 4-AGNR/7-BNNR/4-AGNR heterostructure, contains two 4-AGNRs.
The transmission coefficient of u-AGNR/m-BNNR/b-AGNR structures
favors u-AGNR and b-AGNR with $N_z = 3p + 2 $ widths. For wider
AGNR structures like 17-AGNR and 19-AGNR, such as 17-AGNR/7-BNNR
and 19-AGNR/9-BNNR, their transmission coefficients yield results
similar to those of the 15-AGNR/5-BNNR. If we consider the
narrowest width in w-AGNR/n-BNNR structures, it becomes evident
that the 15-AGNR/5-BNNR structure will exhibit the highest power
factor.

\begin{figure}[h]
\centering
\includegraphics[angle=0,scale=0.3]{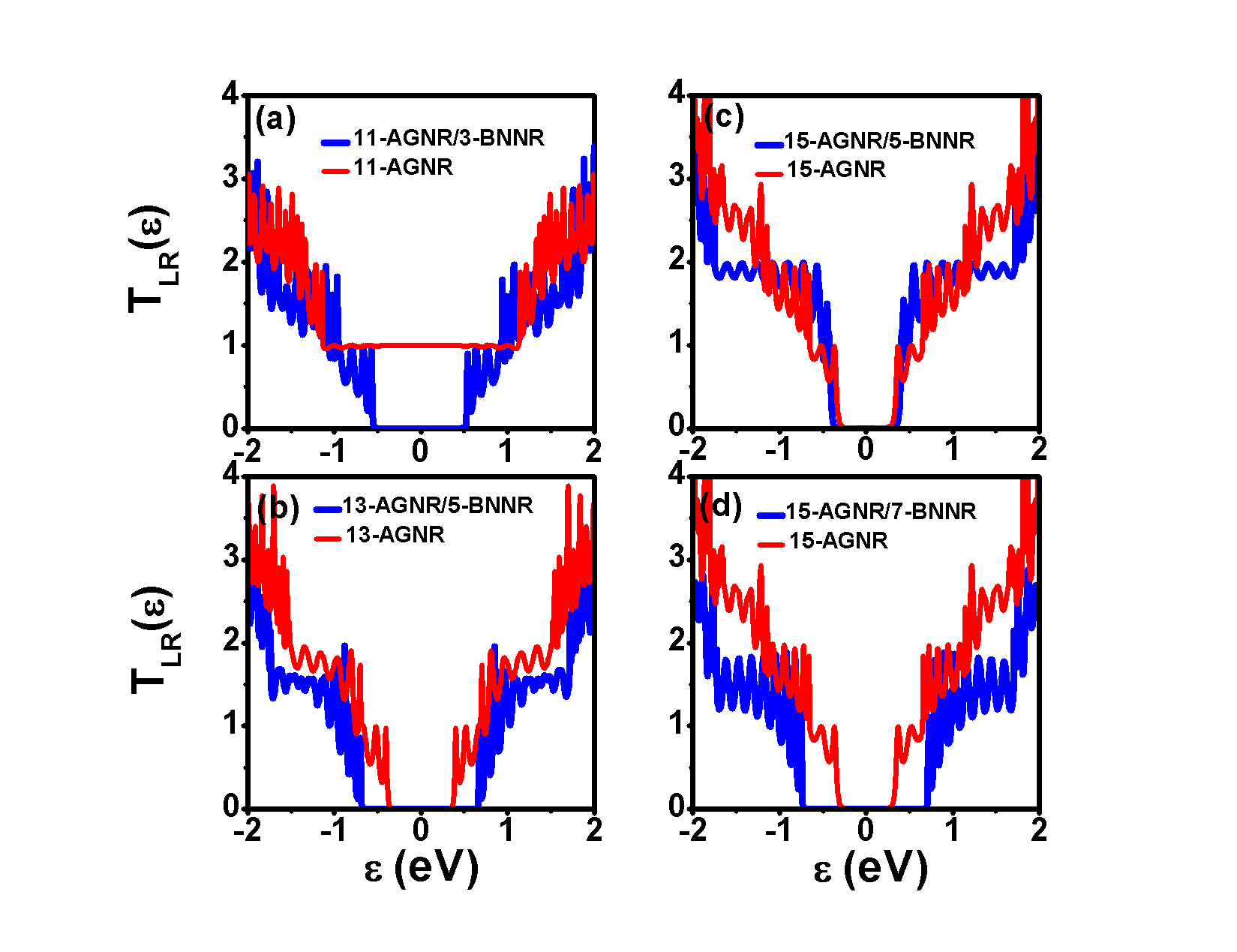}
\caption{Transmission coefficients ${\cal T}_{LR}(\varepsilon)$ of
w-AGNR/n-BNNR structures and w-AGNR structures with $N_a = 84$
($L_a = 8.8$ nm) and $\Gamma_t = 2.7$ eV. (a) 11-AGNR/3-BNNR and
11-AGNR, (b) 13-AGNR/5-BNNR and 13-AGNR, (c) 15-AGNR/5-BNNR and 1D
ideal case, and (d) 15-AGNR/7-BNNR and 15-AGNR.}
\end{figure}

\subsection{Thermoelectric Properties of Finite 15-AGNR/5-BNNR Structures}
In this subsection, we present the calculated electrical
conductance ($G_e$), Seebeck coefficient ($S$), power factor
($PF=S^2G_e$), and figure of merit ($ZT$) for both 15-AGNR/5-BNNR
and 15-AGNR structures as functions of chemical potential at a
temperature of $324$ K and a nanoribbon width of $N_a = 84$ ($L_a
= 8.8$ nm), as depicted in Figure 5. We use specific constants for
the units: $G_0 = 2e^2/h = 77.5~\mu S$ for electrical conductance,
$k_B/e = 86.25 ~\mu V/K$ for the Seebeck coefficient, and
$2k^2_B/h = 0.575~pW/K^2$ for the power factor. In Figure 5(a), we
observe that the electrical conductance of the first conduction
subband exhibits a two-fold quantum conductance value for the
15-AGNR/5-BNNR structure. This significant enhancement in $G_e$
can be attributed to the AGNR array characteristic. In Figure
5(b), the Seebeck coefficient is also enhanced due to the large
band gap. The combined enhancement of $G_e$ and $S$ results in a
substantial maximum power factor value of $PF = 1.326$ in Figure
5(c). It's worth noting that the maximum power factors of AGNRs
with $N_a = 84$ for various ribbon widths ($N_z = 7$, $N_z = 9$,
$N_z = 13$, and $N_z = 15$) are as follows: $0.4457$, $0.7057$,
$0.667$, and $0.803$, respectively. The maximum power factor of
the 15-AGNR/5-BNNR structure not only surpasses these maximum PF
values but also exceeds the theoretical limit, $PF_{QB} = 1.2659$,
as established for one-dimensional ideal systems by Whitney
[\onlinecite{Whitney}].

The thermoelectric figure of merit, denoted as $ZT$, is determined
by the formula $ZT = \frac{S^2G_eT}{\kappa_e + \kappa_{ph}}$,
where $\kappa_{ph}$ represents the phonon thermal conductance of
15-AGNR/5-BNNR structure. For simplicity, we consider $\kappa_{ph}
= F_s
* \kappa_{GNR}$. Here, $\kappa_{GNR}=\frac{\pi^2k^2_BT}{3h}$
denotes the phonon quantum conductance of 15-AGNRs. $F_s = 0.1$
denotes a reduction factor resulting from w-AGNR/n-BNNR
heterostructures. It has been theoretically demonstrated that the
magnitude of $\kappa_{ph}$ can be reduced by one order magnitude
for AGNRs with a BN interface[\onlinecite{TranVT}]. The calculated
$ZT$ values for 15-AGNR/5-BNNR and 15-AGNR structures are
presented in Figure 5(d). Due to the enhanced power factor, the
$ZT$ of 15-AGNR/5-BNNR is increased by $56\%$ compared to that of
15-AGNR, which exhibits $ZT = 1.779$. It should be noted that in
the calculations of $ZT$ values, the phonon thermal conductance
for 15-AGNR/5-BNNR and 15-AGNR structures is assumed to be the
same. However, it is expected that the $\kappa_{ph}$ of
15-AGNR/5-BNNR is smaller than that of 15-AGNR
[\onlinecite{TranVT}]. Consequently, the $56\% $ enhancement in
$ZT$ for 15-AGNR/5-BNNR may be considered a conservative estimate.

\begin{figure}[h]
\centering
\includegraphics[angle=0,scale=0.3]{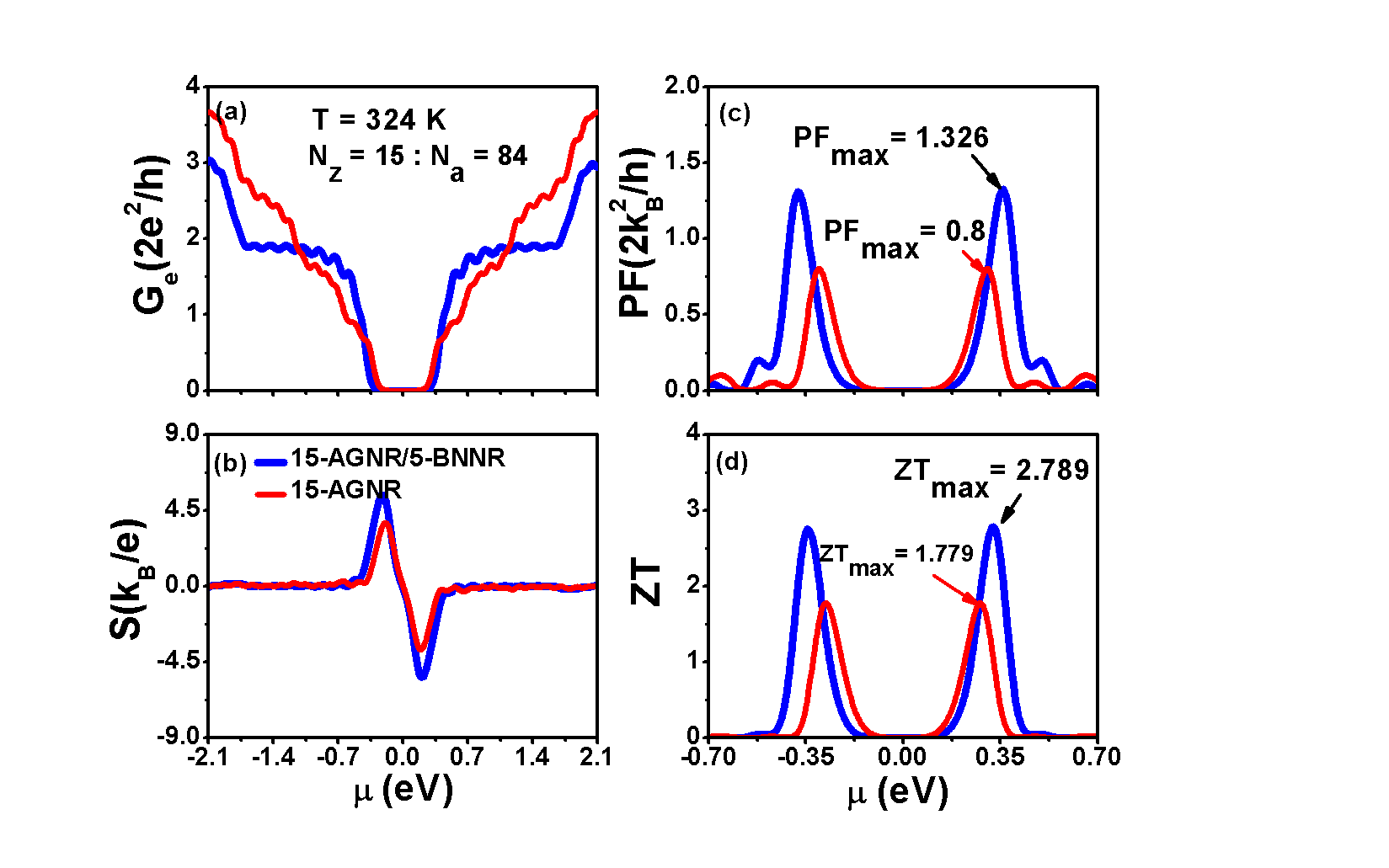}
\caption{(a) Electrical conductance ($G_e$), (b) Seebeck
coefficient ($S$), (c) power factor ($PF = S^2 G_e$), and (d)
figure of merit of 15-AGNR/5-BNNR and 15-AGNR structures with a
ribbon length of $N_a = 84$ ($L_a= 8.8$ nm), as functions of
chemical potential ($\mu$) at a temperature of $324$ K.}
\end{figure}

\section{Conclusion}
In our study, we investigated the electron transport properties
along the armchair edge direction of w-AGNR/n-BNNR. Notably, when
the zigzag edge atoms of these heterostructures make direct
contact with the electrodes, the transmission coefficient area of
the 5-AGNR/5-BNNR/5-AGNR structure surpasses the ideal 1D case, as
illustrated in Figure 4(c). This remarkable outcome can be
attributed to two crucial factors: (a) The electronic properties
of 5-AGNRs undergo a transition from metallic phases to
semiconducting phases when the BNNR potential barrier height
replaces one of the two vacuum potential barrier heights. (b) The
15-AGNR/5-BNNR structure exhibits AGNR array characteristics. The
formal factor leaves the Seebeck coefficient unchanged, and the
latter factor significantly enhances the electrical conductance.
The maximum power factor achieved by the finite
5-AGNR/5-BNNR/5-AGNR structure, with a ribbon length of $N_a = 84$
($L_a = 8.8$ nm), reaches $PF = 1.326~\frac{2k^2_B}{h}$ at a
temperature of $324$ K, surpassing the quantum limit for the power
factor in 1D systems, $PF_{QB} = 1.2659~\frac{2k^2_B}{h}$. Due to
the mechanical strength and flexibility of GNR thermoelectric
generators (TEGs), the 15-AGNR/5-BNNR structures show promise for
utilizing semiconducting GNR-based TEGs in various applications,
including wearable electronics [\onlinecite{WeiTY}]. Our design
not only represents a significant advancement in GNR-based
thermoelectric devices but also holds the potential for enhancing
the performance of GNR-based optoelectronics
[\onlinecite{WangHM}].


{}

{\bf Acknowledgments}\\
This work was supported by the National Science and Technology
Council, Taiwan under Contract No. MOST 107-2112-M-008-023MY2.

\mbox{}\\
E-mail address: mtkuo@ee.ncu.edu.tw\\

 \numberwithin{figure}{section} \numberwithin{equation}{section}

\setcounter{section}{0}
\setcounter{equation}{0} 

\mbox{}\\





\end{document}